\documentclass[aps,showpacs,superscriptaddress,twocolumn]{revtex4}
\usepackage{amsmath}
\usepackage{graphicx}

\begin{document}

\title{Electric Dipole Polarizabilities of Hydrogen and Helium Isotopes}
\author{I. Stetcu}

\altaffiliation{Present address: Department of Physics, University of Washington, Box 351560, Seattle, Washington, 98195-1560.}

\affiliation{Theoretical Division, Los Alamos National Laboratory, 
Los Alamos, New Mexico 87545}

\author{S. Quaglioni}
\affiliation{Lawrence Livermore National Laboratory, Livermore, P.O. Box 808,
California 94551}

\author{J.L. Friar}
\affiliation{Theoretical Division, Los Alamos National Laboratory, 
Los Alamos, New Mexico 87545}

\author{A.C. Hayes}
\affiliation{Theoretical Division, Los Alamos National Laboratory, 
Los Alamos, New Mexico 87545}

\author{Petr Navr\'{a}til}
\affiliation{Lawrence Livermore National Laboratory, Livermore, P.O. Box 808,
California 94551}

\date{\today} 

\begin{abstract}

\hspace*{-1.3in}
\rotatebox{90}{%
\fbox{\parbox[t]{1.15in}{LLNL-JRNL-411284\\
LA-UR-09-01482}}
}
\vspace*{-1.05in}

The electric dipole polarizabilities of $^3$H, $^3$He, and $^4$He are calculated directly using the Schr\"odinger equation with the latest generation of two- and three-nucleon interactions. These polarizabilities are necessary  in order to obtain accurate nuclear-polarization corrections for transitions involving  S-waves in  one- and two-electron atoms.  Our results are compared to previous results, and it is shown that direct calculations of the electric polarizability of $^4$He using modern nuclear potentials are smaller than published values calculated using experimental photoabsorption data. The status of this topic is assessed in the context of precise measurements of transitions in one- and two-electron atoms.

\end{abstract}

\maketitle

\section{Introduction}

Theoretical calculations of transition frequencies in hydrogenic atoms and ions have reached a level of precision where tiny corrections due to nuclear structure and dynamics are necessary in order to interpret the results of high-precision measurements in these systems. This has largely been the result of recent improvements in Quantum Electrodynamics (QED) calculations~\cite{codata2006,savely,EGS}. In many cases the experimental errors and estimated sizes of uncalculated QED corrections are much smaller than the nuclear corrections, and one can thus use those measurements (corrected for QED effects) as an experimental determination of various nuclear quantities~\cite{FMS,savely02}. We briefly review several such determinations.

For S-wave hyperfine transitions in one-electron atoms and ions~\cite{savely02,hyper1,hyper2}, experimental precision is much greater than that of all theoretical calculations, while uncalculated theoretical contributions  to transition frequencies (including QED corrections) are significantly smaller than nuclear effects.  The leading-order (i.e., largest) nuclear contribution to these hyperfine transitions (called a Low moment~\cite{low}) is determined by a correlation between the nuclear charge and  current operators~\cite{hyper1,hyper2}. Low moments may be further decomposed into  Zemach moments~\cite{zemach}  (viz., utilizing only ground-state expectation values of the charge and current operators) and polarization contributions (viz., including only virtual excited states between the two operators), both of which  play significant roles. For the important proton (i.e., $^1$H) case the polarization effects are significantly smaller than the static (Zemach) corrections because the proton is much more difficult to excite than any nucleus~\cite{fs04,ingo06,carl06}. Although exceptionally interesting, hyperfine transitions are not the focus of this paper.

The  frequencies of transitions between S-states in hydrogenic atoms and ions can be separated into  
a reference value (essentially the Dirac transition frequency modified by reduced-mass effects) plus the much smaller Lamb shift. The Lamb shift contribution is dominated by QED corrections, but nuclear effects play a significant role in the best measured transitions. These nuclear corrections can be decomposed into nuclear finite-size corrections (i.e., determined by nuclear ground-state charge distributions)  plus nuclear polarization corrections (viz., involving virtual excited states of the nucleus). The latter are typically dominated and determined by the electric polarizability, which reflects the distortion of the nuclear charge distribution as it  is attracted by (and follows) the orbiting electron.

The most accurate measurement of such a frequency was performed in Ref.~\cite{fischer04} for the 1S-2S transition in hydrogen, with a relative error of slightly more than 1.4 parts in $10^{14}$ and with an absolute error of 34 Hz. That error is slightly smaller than the estimated polarization correction of 60(11) Hz from Ref.~\cite{ks00}, and is much smaller than the size correction of about 1000 kHz. The mismatch in the sizes of these nuclear corrections reflects both their different dimensional structure~\cite{codata2006} and the fact that the proton is difficult to excite (compared to a nucleus), even though the proton size is not significantly smaller than that of light nuclei. If one turns the problem around and  extracts the proton-size correction from the experimental transition frequency~\cite{codata2006}, one obtains a value for the  proton r.m.s. charge radius of $\langle r^{2} \rangle_{ch}^{1/2}$ = 0.877(7) fm, which agrees with a recent direct determination of that quantity from elastic electron-scattering data~\cite{sick03,blunden}: $\langle r^{2} \rangle_{ch}^{1/2}$ = 0.897(18) fm. Both the polarization-correction and experimental errors are much smaller than the Rydberg constant error~\cite{codata2006}, which dominates the uncertainty in the hydrogen atom analysis. 

A similar analysis of transitions from the 2S state in deuterium to a variety of S and D states~\cite{codata2006} leads to a value for the deuteron charge radius of  $\langle r^{2} \rangle_{ch}^{1/2}$ = 2.1402(28) fm, which is consistent with the electron scattering value~\cite{d-sick,sick-d} of  $\langle r^{2} \rangle_{ch}^{1/2}$ = 2.130(10) fm. We note that this is the full charge radius (containing the finite sizes of the proton and neutron constituents), and that the atomic value has an uncertainty nearly 4 times smaller than the value obtained directly from electron scattering.

The determination of the difference in transition frequencies between hydrogen and deuterium for identical transitions can be used to test our understanding of small contributions to the charge radius of the deuteron~\cite{FMS}.  In such a difference nuclear-mass-independent terms (including difficult-to-calculate QED contributions) cancel, which greatly simplifies the analysis. Because the finite size of the proton contributes linearly to the deuteron mean-square radius (which is the nuclear quantity that determines the dominant nuclear-size correction in an atom), it largely cancels out in the frequency difference.  Higher-order proton-size corrections and neutron-size corrections are relatively small and tractable. The transition-frequency difference (dominated by calculable reduced-mass effects)  was reported in Ref.~\cite{huber} for 1S-2S transitions with a relative error of 2.2 parts in $10^{10}$ and an absolute error of 0.15 kHz. The nuclear electric polarizability of deuterium contributes 19.26(6) kHz~\cite{dipole-L}, which is more than two orders of magnitude greater than the experimental error, while the deuteron-size correction is greater than 5000 kHz. The weak binding of the deuteron makes possible the calculation of the bulk of the polarization and nuclear-size corrections in terms of a few well-measured parameters. The tiny remaining size correction includes statistically significant contributions to the nuclear charge radius arising from meson-exchange currents and relativistic corrections~\cite{FMS,stpete-CJP}, which are unobtainable from other types of experiments such as electron scattering. Obtaining this sensitivity to fine details of nuclear dynamics depends on accurate estimates of the deuteron electric polarizability.

Measurements of S-wave transition frequencies in $^3$H, $^3$He, and $^4$He atoms are not yet as accurate as those described above, nor are the necessary theoretical calculations for He atoms. It may be possible to improve~\cite{comb} both to the point where nuclear physics information can be extracted, particularly information about the r.m.s. charge radii. 
As reviewed and updated in Ref.~\cite{drake-rev}, on the other hand, isotopic differences in transition frequencies for helium and singly ionized lithium isotopes now have the required experimental and theoretical sensitivity. The latter sensitivity is greatly enhanced by the cancellation of nuclear-mass-independent relativistic and QED corrections in isotopic differences. In complete analogy to the hydrogen-deuterium case, calculable reduced-mass effects dominate the frequency differences, leaving nuclear contributions as the residue after relativistic and QED contributions are subtracted. There has been considerable recent interest in the isotope shifts of $^3$He~\cite{shiner,he3-drake}, $^6$He~\cite{argonne,DY,argonne-II}, and $^8$He~\cite{argonne-II} transitions relative to those of $^4$He. In each case the value of the r.m.s. charge radius of that isotope has been extracted relative to the charge radius of $^4$He~\cite{BR,sick2008}. The nuclear polarizability correction to the $^3$He - $^4$He isotope-shift frequency (the best measured of the He isotope shifts) is about 2/3 of the 3 kHz experimental uncertainty~\cite{drake-rev,he3-drake,shiner}, while presently only a marginal influence~\cite{drake-rev} on the others, but future improvements should require reliable values of their electric polarizabilities (as was the case for the deuteron), and that is the purpose of this paper.

\section{Calculational Techniques}

The electric (dipole) polarizability of a nucleus (or atom),  $\alpha_{\rm E}$, is defined by 
\begin{equation}
\alpha_{\rm E} = 2 \alpha \; \sum_{N \neq 0} \; \frac{| \langle N |  D_z | 0 \rangle |^2}{E_N - E_0} \, ,
\label{alphaE_def}
\end{equation}
where $\alpha$ is the fine-structure constant, $E_0$ is the energy of the ground-state $|0\rangle$, $E_N$ is the 
energy of the N${\underline{th}}$ excited state, $|N\rangle$ (all of which are in the continuum for few-nucleon systems), and $D_z$ is the component  of the (non-relativistic, in our case) electric-dipole operator in the $\hat{z}$ direction, which generates the transition between those  states.  The definition (1) can be rearranged into the form of a sum rule  
\begin{equation}
\alpha_{\rm E} = \frac{1}{2 \pi^2} \int_{\omega_{\rm th}}^{\infty} d \omega \frac{\sigma^{\rm ud}_{\gamma}
(\omega)}{\omega^2} \, \equiv \frac{\sigma_{-2}}{2 \pi^2}\, ,
\label{alphaE}
\end{equation}
where $\sigma^{\rm ud}_{\gamma} (\omega)$ is the nuclear cross section for  
photoabsorption of unretarded-dipole (long-wavelength) photons with energy $\omega$, and $\omega_{\rm th}$ is the threshold energy for photoabsorption. The inverse-energy weightings in Eqns.~(\ref{alphaE_def}) and (\ref{alphaE}) lead to significant sensitivity of $\alpha_{\rm E}$ to the threshold energy, $\omega_{\rm th}$, which depends on nuclear binding energies.

In order to obtain the nuclear energy spectra and the wave functions involved in the calculation of the electric polarizability (Eqn.~(\ref{alphaE_def})), we use
the no-core shell model (NCSM) in relative coordinates~\cite{Navratil:1999pw} to solve the Schr\"odinger equation. The NCSM is a flexible approach to solving the few- and many-nucleon problems, and it has been extensively used in studies of $s$- and $p$-shell nuclei~\cite{Navratil:2001,Hayes:2003ni,Navratil:2007,Navratil_FBS,Quaglioni:2007eg}. In the NCSM the nuclear wave functions are obtained by the diagonalization of an effective Hamiltonian in a finite basis constructed from harmonic oscillator (HO) wave functions. The truncation of the model space is taken into account via an effective interaction derived by means of a unitary transformation. Either local or non-local nucleon-nucleon (NN) and three-nucleon (NNN) interactions can be used in the Schr\"odinger equation. 
The effective interaction is constructed in a cluster approximation, which must be truncated for practical reasons. The truncation of the model space is determined and labeled by the number of excitations, $N_{max}$, above the non-interacting state. We test convergence by plotting calculated quantities vs. $N_{max}$, and those quantities should approach their correct asymptotic values as $N_{max}$ becomes infinite. Thus by observing the convergence of observables as a function of $N_{max}$, we can determine their values.

In this paper we compute the $^3$H, $^3$He, and $^4$He electric dipole polarizabilities starting from a 
nuclear Hamiltonian derived within the framework of (QCD-based) Chiral Perturbation Theory (including 
the Coulomb interaction between the protons). We adopt the nucleon-nucleon interaction at next-to-next-to-next-to-leading order (or N$^3$LO) of Ref.~\cite{N3LO} and the three-nucleon interaction at next-to-next-to-leading order (or N$^2$LO)~\cite{vanKolck:1994,Epelbaum:2002} in the local form of Ref.~\cite{Navratil_FBS}. The accuracy of these nuclear interactions for $s$- and $p$-shell nuclei was investigated extensively in the same NCSM framework in Ref.~\cite{Navratil:2007}. In particular, the experimental binding energies of $^3$H and $^3$He are reproduced with high accuracy (viz., within 8 keV, or about one part per thousand)~\cite{Navratil:2007,Navratil_FBS,gazit-2008}, while the $^4$He ground-state energy is within a few hundred keV of experiment (i.e., within approximately 1\%). This residual discrepancy with experiment reflects our current uncertainty on the underlying nuclear dynamics.  These modern nuclear 
forces therefore provide an accurate description of the structure of the nuclides considered here ($^3$H, 
$^3$He, and $^4$He) as well as the total photoabsorption cross section of $^4$He~\cite{Quaglioni:2007eg} (discussed in the next section).

For each nucleus we first solve the few-nucleon Schr\"od§inger equation in order to obtain the ground-state wave function, which can be accurately described in a large HO basis. We next rearrange 
Eqn.~(\ref{alphaE_def}) according to Podolsky's technique~\cite{podolsky}, which allows the ground state to be used as the driving term for the Lanczos-moment method~\cite{LanczAlg1950,haxton:065501}, which is our numerical method of choice for solving the Schr\"odinger equation.  This trick allows us to work only with bound-state quantities and to bypass the much more difficult approach of computing a response in the continuum.  A detailed description of this method in a NCSM framework was given in Ref.~\cite{Stetcu:2008vt}.

\section{Results of Calculations} 

\begin{figure}[t]
\centering\includegraphics[clip,scale=0.52]{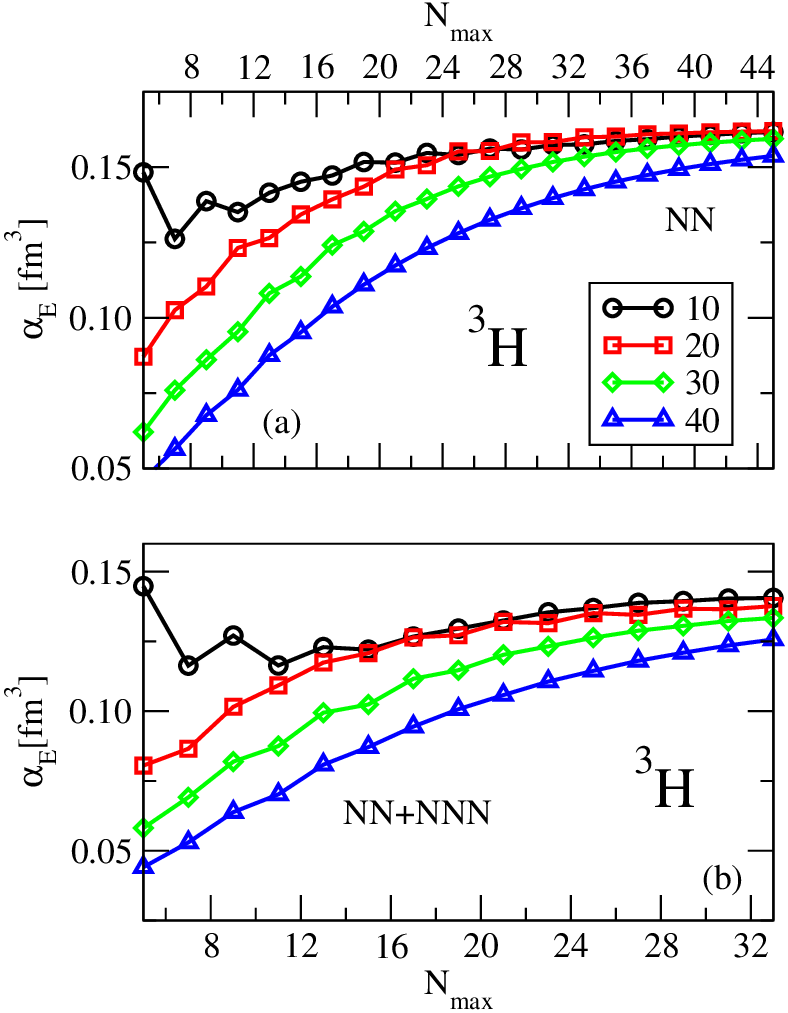}
\caption{[Color online] The dependence of (the electric polarizability) $\alpha_{\rm E}$ of $^3$H (in units of fm$^3$) on the model-space truncation parameter, $N_{max}$. The results have been obtained using (a) NN  interactions only, and (b) NN+NNN interactions. Each curve is obtained using a different frequency parameter for the basis states, shown in the legend in MeV. For sufficiently large $N_{max}$ each result should be independent of that frequency.}
\label{fig:polH} 
\end{figure}

\begin{figure}
\centering\includegraphics[clip,scale=0.52]{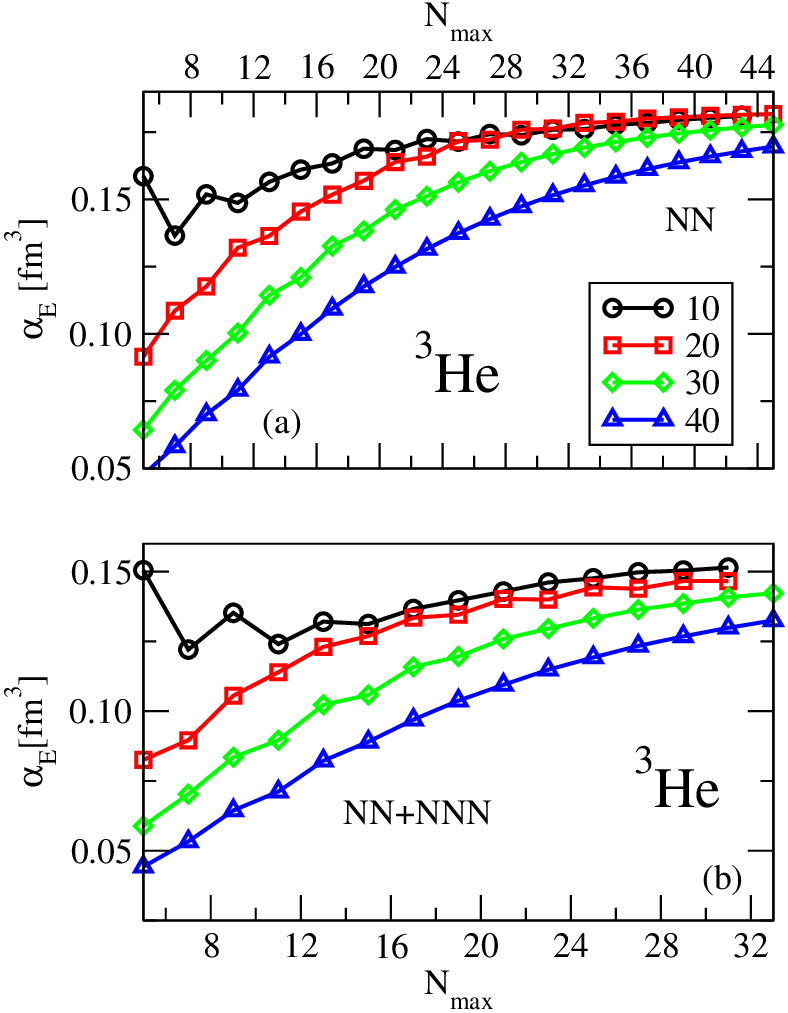}
\caption{[Color online] Same as in Fig. \ref{fig:polH}, but for $^3$He.}
\label{fig:pol3He} 
\end{figure}

\begin{figure}[htbp]
\centering\includegraphics[clip,scale=0.42]{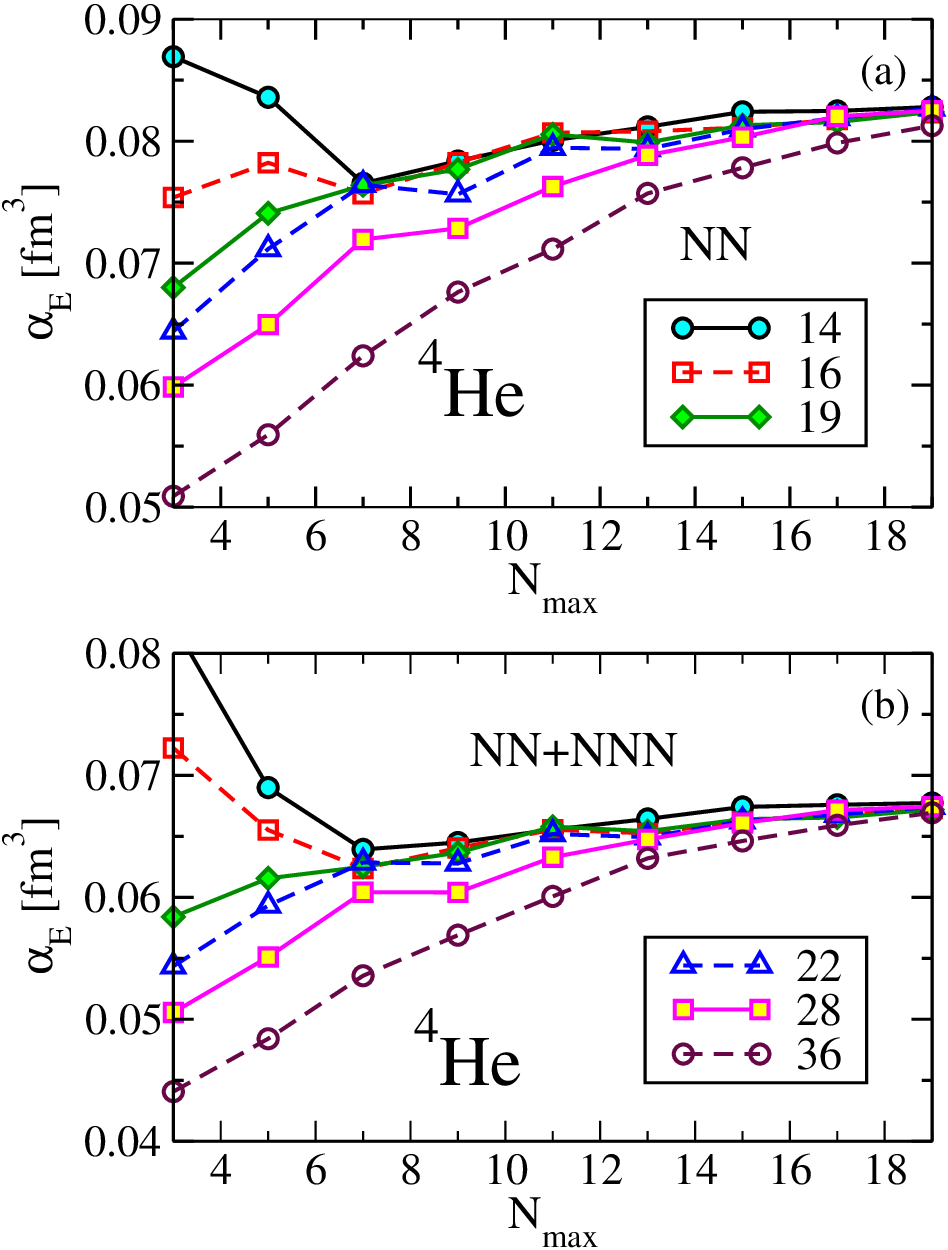}
\caption{[Color online] Same as in Fig. \ref{fig:polH}, but for $^4$He.}
\label{fig:pol4He} 
\end{figure}

In Figs. \ref{fig:polH}--\ref{fig:pol4He}, we show the running of the electric polarizability with the truncation parameter for the model space, $N_{max}$. 
Different HO frequency parameters, $\Omega$, result in different convergence patterns for the electric polarizability, and this fact is especially visible for small $N_{\rm max}$ values. As shown in Figs. \ref{fig:polH}--\ref{fig:pol4He}, results obtained using smaller HO frequencies (equivalent to  larger-length oscillator parameters, defined by $b=1/\sqrt{m_N \Omega}$) converge faster. Long-range operators (such as the electric dipole 
operator) are thus better described using smaller values of $\Omega$ in the smaller model spaces. Moreover, better sampling of the low-lying excited states (the most important states for the calculation of the electric polarizability) is obtained for small values of $\Omega$. Although not shown, other operators converge faster for larger HO frequencies. Binding energies, for example, achieve the fastest convergence for a HO length parameter $b$ on the order of the size of the nucleus considered. For each observable the results    that are obtained with different values of $\Omega$ nevertheless approach a single asymptotic value for large $N_{max}$. Uncertainties in determining that asymptotic value lead to  error estimates in Table~\ref{table:alphaE}.

The upper panels of Figs.~\ref{fig:polH}--\ref{fig:pol4He} present results obtained by neglecting three-nucleon interactions, while results that include three-nucleon interactions are shown in the lower panels. We note, however, that because binding energies are significantly smaller than experiment in the absence of  three-nucleon interactions, the values obtained with only NN interactions are about 10--25\% larger than the results obtained when NNN interactions are included. This is partly the effect of having too small a value for $\omega_{\rm th}$, which emphasizes smaller values of the energy denominators in Eqns.(\ref{alphaE_def}) and (\ref{alphaE}). 

The stronger binding of $^4$He than that of the three-nucleon systems and our slightly less accurate reproduction of the $^4$He experimental binding energy may affect our results for $\alpha_{\rm E}$.  In order to probe this possibility  we have performed three calculations with slightly different parameter sets in the three-nucleon force. The specific results are given in~\cite{comment}.

We expect from Eqn.~(\ref{alphaE}) that $\alpha_{\rm E}$ should scale dimensionally like the square of the nuclear size divided by the binding energy. Moreover the square of the size should scale roughly like the inverse of the binding energy (this statement is highly accurate for the weakly bound deuteron).  The resulting dependence on the binding energy should be roughly like the inverse square, and our slight over-binding could diminish  $\alpha_{\rm E}$ by as much as 2\%. Our calculation that neglects the  three-nucleon force in $^4$He results in a value of  $\alpha_{\rm E} = 0.0822(5)$ fm$^3$ (22\% higher than a result of 0.0673(5) fm$^3$ that incorporates this force), while reducing the binding energy from 28.50(3) to 25.39(1) MeV (an 11\% decrease). A similar effect is also seen in the $^3$He and $^3$H calculations.

A more difficult problem is that stronger binding emphasizes nuclear corrections of relativistic order, including corrections to the electric-dipole operator from meson-exchange currents, which are determined by details of how the nuclear forces are constructed~\cite{forward}. This effect could be as large as $\pm 2\%$ for the well-bound $^4$He, but has been little studied and takes us far beyond the scope of this work.We will incorporate these uncertainties into our results in the next section.

\section{Comparison with other work}

Our calculations of the electric polarizabilities of three- and four-nucleon isotopes of hydrogen and helium  are summarized in Table~\ref{table:alphaE}, together with  those of others using different two-nucleon and three-nucleon forces.  We have restricted our own entries to those that incorporate three-nucleon forces and hence have accurate binding energies, especially for the three-nucleon systems and slightly less so for $^4$He.  For completeness  in the table we have also included the deuterium and $^6$He cases, which were not treated in this work.  

\begin{table}[t]
\caption{Values of the electric polarizability of light nuclei, both theoretical
and experimental, in units of fm$^3$.  The experimental results have been determined by nuclear experiments, including the use of experimental photoabsorption data in Eqn.~(\ref{alphaE}).
No uncertainties were given for the $^3$H, $^3$He, and $^4$He calculations in
\protect~\cite{efros,winfried}, but they are likely to be smaller than about 10\%.  The $^6$He result is
a hybrid calculation relying on some theoretical input and we add it here for completeness.
Results from the present calculation have no listed reference. Our three separate results\cite{comment} for $^4$He have been combined in the table and are discussed in the text near the end of this section. The result of Ref.~\cite{phillips} for the deuteron is an Effective Field Theory calculation. The errors for the two deuteron calculations are not independent and should not be combined. 
\label{table:alphaE}}
\begin{ruledtabular}
\begin{tabular}{lllll}
Nucleus & $\alpha_{\rm E}^{\rm calc}\, ({\rm fm}^3)$ \hspace{0.2in}
&ref. \hspace{0.2in}
&$\alpha_{\rm E}^{\rm exp}\, ({\rm fm}^3)$ \hspace{0.2in}
&ref. \\ \noalign{\smallskip} \hline
$^2$H  & 0.6328(17)&~\cite{dipole-L}  & 0.61(4)   &~\cite{e-d} \\
       &  0.6314(19) &~\cite{phillips}     & 0.70(5)   &~\cite{rodning} \\ \hline
$^3$H  & 0.139(2)    & &\;\;\;\mbox{$-$}   & \rule{0in}{2.5ex} \\
       &         0.139   &~\cite{efros} & \\ \hline
$^3$He & 0.149(5) &  &0.250(40) &~\cite{he3-pol}  \\
       & 0.145      &~\cite{winfried}   & 0.130(13) &~\cite{rinker} \\
       &  0.153(15)          & ~\cite{pachucki}            &  & \\ \hline
$^4$He & 0.0683(8)(14) &  & 0.072(4)  &~\cite{He4-pol} \rule{0in}{2.5ex}\\ 
       & 0.0655(4)     &~\cite{Gazit2006} & 0.076(8) &~\cite{pachucki}  \\ 
       & 0.076      &~\cite{winfried}   &         &  \\
\hline
 $^6$He & &  & 1.99(40) &~\cite{pachucki}
\end{tabular}
\end{ruledtabular}
\end{table}

Only one other calculation of $\alpha_{\rm E}$ for $^3$H exists~\cite{efros}, and our result is in agreement with that calculation.

Calculations for the electric polarizability of $^3$He~\cite{winfried, pachucki} are in agreement within their uncertainties, and are in reasonable agreement with the determination of Ref.~\cite{rinker}, but not with Ref.~\cite{he3-pol}. We note that if charge symmetry were exact in the three-nucleon systems, the Hamiltonians and polarizabilities of $^3$H  and $^3$He  would be identical. Under the charge-symmetry operation that relates the two nuclei the dipole operators in Eqn.~(\ref{alphaE_def}) would each develop a minus sign, while the radial wave functions and Green's functions would be identical. Most of the charge-symmetry violation in these systems is caused by the repulsive Coulomb interaction between the two protons in $^3$He. We note that our uncertainties for these two nuclei are also different. The repulsive Coulomb interaction in $^3$He leads to a larger radius for that nucleus, and that may be the source of the larger uncertainty. Matrix elements of infrared operators (i.e., those like the dipole operator that are most sensitive to large distances from the center of a nucleus) converge more slowly in the NCSM than do short-range operators, which can be successfully renormalized~\cite{Stetcu:2004wh,Stetcu:2006zn}.

The uncertainty in the underlying nuclear dynamics (rather than the uncertainties reflected in the convergence plots) dominates the error estimate of our calculated electric polarizability of $^4$He (see~\cite{comment}). After correcting our three values (corresponding to three nuclear force models) for overbinding~\cite{comment} we average them and use their spread as our direct uncertainty, with an additional 2\% systematic uncertainty from the nuclear dynamics (discussed in Section III). This produces $\alpha_{\rm E} =  0.0683(8)(14) $ fm$^3$, which is listed in  Table~\ref{table:alphaE}. Our result there is significantly smaller than most of the corresponding results, although just at the limit of the estimated uncertainties. We are, however, in fairly good agreement with a recent calculation by Gazit {\it et al.}~\cite{Gazit2006}, which predicts a slightly smaller polarizability, but also corresponds to a slightly overbound model. We note that Ref.~\cite{winfried} used a very primitive nuclear force model and that those results are superseded by those of Ref.~\cite{Gazit2006}. References~\cite{pachucki} and~\cite{He4-pol} used fits to experimental photoabsorption data and Eqn.~(\ref{alphaE}) in order to obtain their results. Values obtained from a direct solution of the Schr\"odinger equation are therefore at some variance with those calculated using experimental $^4$He photoabsorption data.

Measurements of $^4$He photoabsorption in the near-threshold region have been controversial over the years, particularly with respect to the height of the cross section at the peak, for which one can find differences of up to a factor of two between different experiments (e.g., see Ref.~\cite{Quaglioni:2007eg} and references therein). 
This makes it very challenging to extract an accurate and unambiguous value of the $^4$He electric polarizability from the measured 
$^4$He photoabsorption cross section using Eqn.~(\ref{alphaE}). 
In contrast there has been substantial recent progress in theoretical calculations of the $^4$He photoabsorption cross section. 
Predictions obtained using high precision NN and NNN interaction models (including the ones used in this work) all lie in a rather constrained band~\cite{Quaglioni:2007eg}, in remarkable contrast to the large discrepancies present among  different experimental data. 
This gives us confidence that our prediction for the $^4$He electric polarizability, obtained by direct solution of the Schr\"odinger equation,  will prove to be more accurate than those obtained using existing photoabsorption data.

\section{Conclusion}

We have used the latest generation NN and NNN interactions in a NCSM framework in order to obtain accurate three- and four-nucleon solutions of the Schr\"odinger equation. Using the Lanczos-moment method to implement Podolsky's technique~\cite{podolsky} for treating second-order perturbation theory, we have calculated the electric polarizabilities of $^3$H, $^3$He, and $^4$He. Our result for  $^3$H is in excellent agreement with that of Ref.~\cite{efros}, while that for  $^3$He is in good agreement with previous work.
Our best estimate of 0.0683(8)(14) fm$^3$ for $^4$He based on direct solutions of the Schr\"odinger equation is at the lowest end of the calculations that used experimental photoabsorption data directly in Eqn.~(\ref{alphaE}). 
Future calculations for other light nuclei such as $^6$He and $^6$Li should be tractable, but would require a change of basis for the NCSM.
For nuclei with mass numbers greater than five, a Slater Determinant basis is much more efficient than the relative coordinate approach used in this work.

\acknowledgments
We thank D. Phillips for discussions regarding the polarizability of $^2$H and D. Gazit for providing us with the theoretical uncertainty estimate for
the result in Ref.~\cite{Gazit2006}.
The work of I.S., J.L.F. and A.C.H. was performed under the auspices of the U. S.  
DOE. Prepared by LLNL under contract No. DE-AC52-07NA27344. Support from the 
U.S. DOE/SC/NP (Work Proposal No. SCW0498), and from the U. S. Department of Energy Grant 
DE-FC02-07ER41457 is acknowledged. S.Q. and P.N. thank the Institute for Nuclear Theory for its hospitality and the Department of Energy for partial support during the completion of this work.


\end{document}